\documentclass[a4paper, 11pt]{article} 

\usepackage{hyperref}
\usepackage[english,portuguese]{babel}
\usepackage[utf8]{inputenc}
\usepackage{float}

\usepackage[protrusion=true,expansion=true]{microtype} 
\usepackage{graphicx} 
\usepackage{wrapfig} 
\usepackage{tocloft}

\usepackage{mathpazo} 
\usepackage[T1]{fontenc} 
\usepackage{etoolbox}
\makeatletter
\renewcommand\@biblabel[1]{\textbf{#1.}} 
\renewcommand{\@listI}{\itemsep=0pt} 
\pretocmd{\chapter}{\addtocontents{toc}{\protect\addvspace{5\p@}}}{}{}
\pretocmd{\section}{\addtocontents{toc}{\protect\vspace{-4mm}}}{}{}
\renewcommand{\maketitle}{ 
\begin{flushright} 
{\LARGE\@title} 

\vspace{50pt} 

{\large\@author} 
\\\@date 

\vspace{40pt} 
\end{flushright}
}

\title{\textbf{Ensaio sobre o Auto-Aproveitamento}\\ 
um relato de investidas naturais na participação social} 

\author{\textsc{Renato Fabbri} 
\\{\textit{IFSC/USP, Participa.br/SG-PR, labMacambira.sf.net}}} 

\date{\today} 

\begin{document}

\maketitle 



{
\selectlanguage{english}
\begin{abstract}
The use of digital traces of our social structures and activities is a reality for some companies and State instances. The exploitation by the individual and by Society is still incipient. This writing is a brief account of an immersion to advance this civil empowerment, beginning with experiments for collection and dissemination of information, and going through social structures streaming, resource recommendation via complex networks and natural language processing, linked data and ontological organizations of social and participatory structures.
\end{abstract}
}

\begin{abstract}
O aproveitamento de nossos rastros digitais de estruturas e atividades sociais é uma realidade para algumas empresas e instâncias do Estado. O aproveitamento pelo indivídulo e pela Sociedade ainda é incipiente. Este escrito é um breve relato de uma imersão para adiantar este empoderamento civil, começando por experimentos de coleta e difusão de informação, passando por \emph{streaming} de estruturas sociais, recomendação de recursos via redes complexas e processamento de linguagem natural, dados ligados e organizações ontológicas de estruturas sociais e participativas.
\end{abstract}
%
%
%

\hspace*{3,6mm}\textit{Keywords:} redes complexas, processamento de linguagem natural, dados ligados, participação social, física antropológica 


\newpage
\tableofcontents

\section*{Abertura no fim do mundo}
\addcontentsline{toc}{section}{Abertura no fim do mundo}

Ao final de 2012, foi-me proposto pelo grupo Mutgamb/Metareciclagem, e pelos parceiros Glerm Soares e Simone Bittencourt, a participação  em um trabalho sobre o fim do mundo. Esta ocasião mostrou-se propícia para experimentos em rede, com o propósito de difundir uma prática transformadora, capaz de modificar substancialmente a nossa realidade, de forma a manifestar um ``fim do mundo''. A prática difundida nas redes era sobre o aproveitamento das próprias redes sociais, com ferramentais de redes complexas e processamento de linguagem natural. Os ciclos de difusão são por vezes chamados de ``vaca do fim do mundo'' e relacionados com as passeatas de junho de 2013, pois eram previstas como uma das hipóteses iniciais deste experimento percolatório de \emph{física antropológica}. Estes primeiros momentos desta pesquisa assegurou fertilização por diversos atores de uma rede de amizades com ao menos uma década de explícito ativismo nas áreas de software e mídia livre. Diversos escritos, galerias de imagens e páginas são remanescentes deste primeiro momento~\cite{ciberiun,ars,rc1,rc2}, assim como articulações e amadurecimentos que desembocaram no objeto deste ensaio.
\begin{figure}[H]
  \centering
    \includegraphics[width=.9\textwidth]{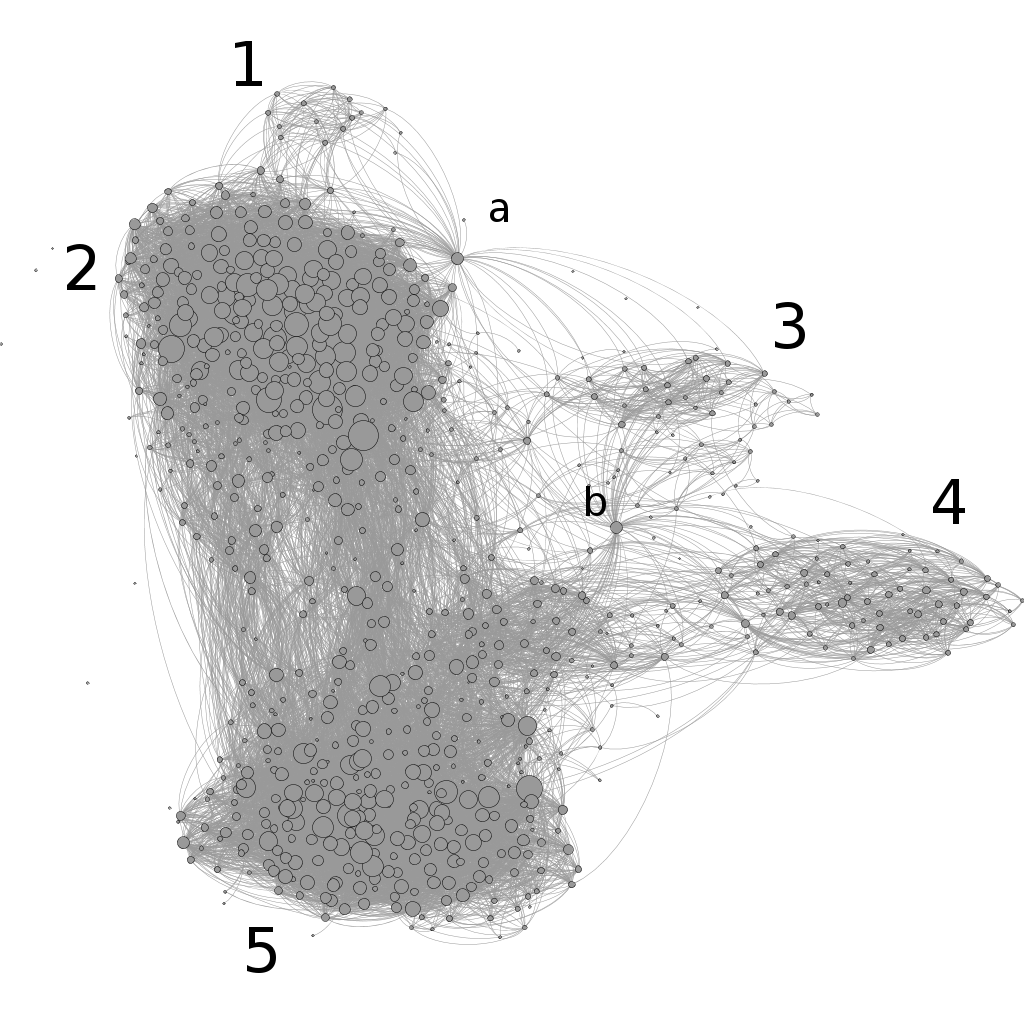}
  \caption{\small Minha rede de amizades do Facebook em Fevereiro de 2012: cada vértice corresponde a um amigo, cada aresta a uma amizade entre os amigos correspondentes aos vértices nas extremidades. Em 1 estão os amigos da igreja da esposa; em 2, amigos de graduação e relacionados; em 3, família por parte de pai e mãe; em 4, conhecidos do IFSC/USP; em 5, conhecidos através do trabalho com tecnologias sociais, artísticas e livres. Em `a' está minha esposa, em `b' meu irmão. Na ``vaca do fim do mundo'', esta rede foi estimulada com 3 ciclos em que cada membro era acionado individualmente, dos menos conectados aos mais conectados.}\label{fig:redem}
\end{figure}

Ao final dos ciclos de coleta e difusão de informação, minha rede (meu eu-rede, ilustrada na Figura~\ref{fig:redem}) havia se rearranjado para acolher o trabalho proposto. Dada a pertinência do assunto e dedicação à absorção e geração de materiais, meu doutorado na física computacional (IFSC/USP) foi alinhado e em simbióse foi aproximado suporte das estruturas federal (SNAS/SGPR) e internacional (PNUD/ONU). Era dezembro/2013 e o ano até o dia de hoje possibilitou o discurso que segue.

\section*{Streaming de estruturas sociais}
\addcontentsline{toc}{section}{Streaming de estruturas sociais}

Uma traço forte que pude observar com os ciclos de coleta e difusão de informação é a apropriação tímida que as pessoas apresentam de suas estruturas sociais. Algumas pessoas se encantam com as figuras, outras com os conceitos e com a percepção do aspecto social de si, por vezes chamado de ``ser-rede'' ou ``eu-rede'' por interessados durante as difusões. Nestes contextos, teorias academicas, como TAR (Teoria Ator Rede, Latour), raramente eram puxadas diretamente, mesmo por especialistas. A eficiência do discurso intuitivo para comunicar sobre os interesses é impressionante. O impulso em relatar as impressões assemelha-se ao impulso de relatar sonhos, com vislumbres de estruturas sutis e inconscientes e sequencial esquecimento.

\begin{figure}[H]
  \centering
    \includegraphics[width=.7\textwidth]{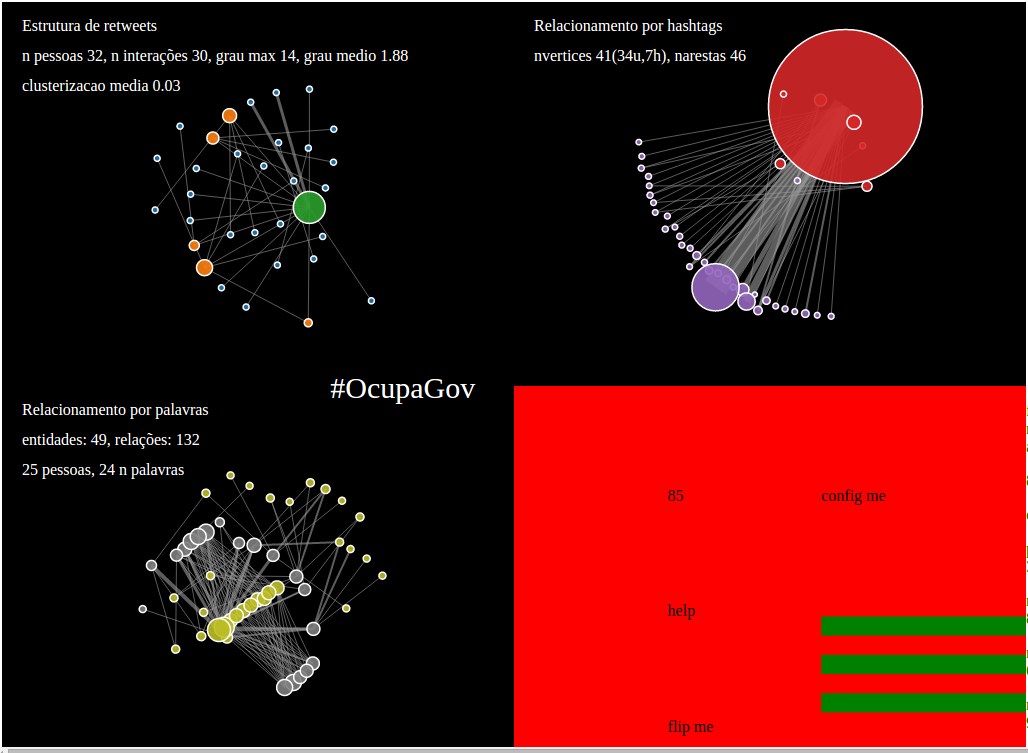}
  \caption{\small Telão para streaming de estruturas sociais, usado no \#arenaNETmundial, \#ocupaGOV e outras ocasiões. Tela com rede de retweets e relacionamento via hashtag e vocabulário. Atualizada a cada 10 segundos com os relacionamentos implicados pelos dos tweets mais recentes.}\label{fig:telao1}
\end{figure}

\begin{figure}[H]
  \centering
    \includegraphics[width=.7\textwidth]{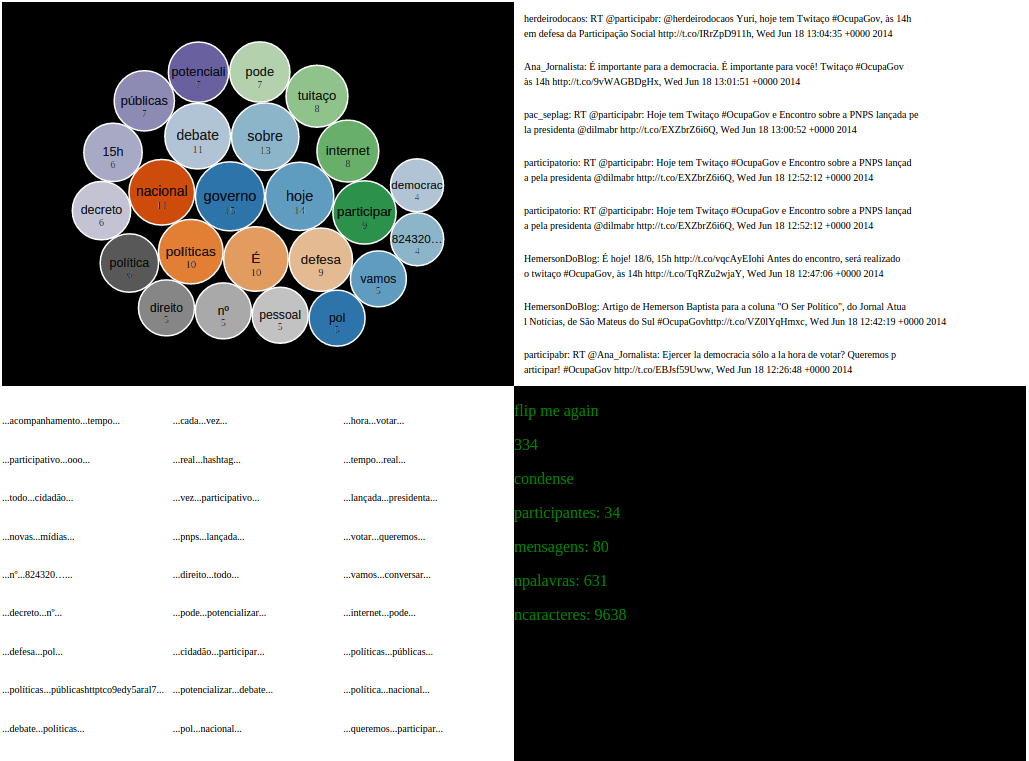}
  \caption{\small Telão para streaming de estruturas sociais, usado no \#arenaNETmundial, \#ocupaGOV e outras ocasiões. Tela com relacionamentos de hashtags e vocabulário. Atualizada a cada 10 segundos com conteúdo dos tweets mais recentes.}\label{fig:telao2}
\end{figure}

Neste contexto, para difundir sobre como os nossos rastros podem ser observados e aproveitados, foram feitos telões de streaming de estruturas sociais em páginas HTML confome a Figura ~\ref{fig:telao1}. Apenas o Twitter foi utilizado, e as redes de retweet e de relacionamento via vocabulário e hashtags eram contempladas. O telão também podia exibir tweets recentes, palavras mais ocorrentes, co-ocorrentes, e outras informações simples de texto, conforme a Figura~\ref{fig:telao2}.


De início, os telões ficaram operantes durante horas no evento \#arenaNETmundial, isso se manteve por $\approx 3$ dias. Eram atualizados a cada 10 segundos com os tweets mais recentes que possuiam as hashtags acompanhadas pelo evento. Instâncias online são~\cite{ocupagov,mmissa,mmissa2}, o código está em~\cite{codTelao}, integrado à uma instância mais ampla~\cite{mmissa2}. Diversas trocas com comunicadores e programadores ajudaram a entender os potenciais de transparência e de conscientização das nossas estruturas em rede. Alguns experimentos a mais foram feitos com streaming de dados, em especial o MyNSA~\cite{mynsa}, parte do conjunto de ferramentas idealizadas para o AARS (Análise e Ação em Redes Sociais).

Ainda assim, os aproveitamentos de nossas estruturas em rede são os mais diversos, apontando para uma multiplicidade de métodos e aplicações. A reflexão por fim nos levou aos mecanimos de recomendação e navegação de recursos.

\section*{Recomendação de recursos e navegação}
\addcontentsline{toc}{section}{Recomendação de recursos e navegação}
A solução encontrada foi a de providenciar métodos diferentes de recomendação de recursos para usos diferentes. As características ideais deste sistema de recomendação, que é um enriquecimento da navegação semântica de recursos (dados ligados/linkados, desenvolvido abaixo), são:

\begin{itemize}
    \item Utilização de quaisquer recursos de referência (artigos, comentários, perfis de usuários) para recomendar outros recursos.
    \item Recomendação por similaridade e dissimilaridade, para os diferentes usos. Por exemplo: pessoas proximas nas redes e com vocabulário similar são potenciais amigos. Pessoas distantes na rede e/ou com vocabulários díspares são potencialmente de outros nichos, opositores políticos ou ideológicos ou pontes para expansão de mobilizações.
    \item Uso de critérios de redes complexas, provenientes ao menos de redes de amizade e de interação entre os participantes.
    \item Uso de critérios linguísticos, provenientes dos conteúdos textuais, ao menos \emph{Bag of Words}.
    \item Explicitação dos critérios usados para cada recomendação.
    \item Sugestão de aproveitamentos para o método usado e para os resultados obtidos (outliers, quantidade, etc).
    \item Disponibilização de interface para testes do usuário com o algorítmo usado em cada recomendação.
    \item Recomendações sob demanda: o usuário requisita que tipo de recursos quer recomendado a partir de que recurso de referência, via quais métodos preferir. Os resultados e estruturas auxiliares são armazenados para otimizar recomendações posteriores e como sugestão para outros usuários.
    \item Navegação ativa, nas quais o usuário faz anotações, ativa recomendações e registra navegação para otimizar usos posteriores.
    \item Permissibilidade para parametrização pelo usuário, transparência nos resultados da recomendação.
\end{itemize}
\begin{figure}[H]
  \centering
    \includegraphics[width=.7\textwidth]{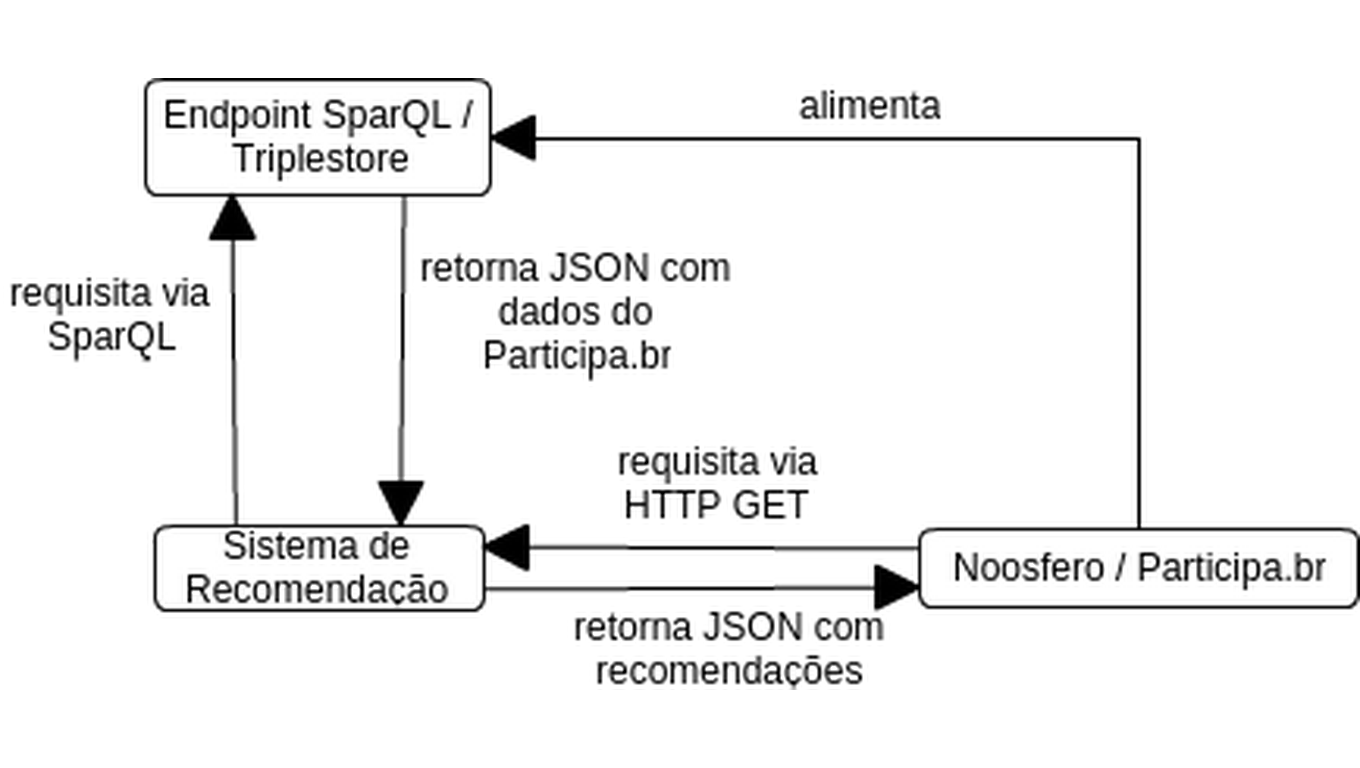}
  \caption{\small Uma das concepções de implementação do sistema de recomendação e suas relações principais com o frontend (Participa.br) e base de dados semanticamente enriquecidos (endpoint SparQL fuseki/jena).}\label{fig:rec}
\end{figure}

Nestes termos, foi feito um sistema de recomendação de participantes e de recursos~\cite{pnud4}. Análises informativas de recursos e tipos de recursos são parte deste sistema de recomendação/priorização de recursos~\cite{pnud3}. Estes métodos de análise, baseados nas estruturas em rede e no processamento de texto, não estão explorados neste ensaio mas são idealizados para realizar análises da nuvem participativa via critérios intuitivos e úteis.
 Uma das implementações concebidas deste sistema de recomendação, para funcionar junto ao Noosfero/Participa.br, está delineada na Figura~\ref{fig:rec}.

\section*{Dados ligados e ontologias}
\addcontentsline{toc}{section}{Dados ligados e ontologias}
A associação de recursos via recomendações amplia associações via critérios ontológicos. Nesta direção, foi revisado o VCPS (Vocabuário Comum de Participação Social), dando origem à OPS (Ontologia de Participação Social)~\cite{OPS}. A OPa (Ontologia do Participa.br) foi levantada com os conceitos da equipe do Participa.br~\cite{OPA} e recebeu um módulo posterior, dedicado aos dados do Participa.br~\cite{pnud5}. Foi feita a OCD (Ontologia do Cidade Democrática), com um método dedicado aos dados, e a OntologiAA (Ontologia do AA), relacionado classes e propriedades a conceitos mais gerais via \texttt{rdfs:subClassOf} e \texttt{rdfs:subPropertyOf} para testes com inferências com o Jena/Fuseki. Foram feitas rotinas para representação em RDF dos dados do Participa.br, Cidade Democrática e AA (rotinas de triplificação de dados)~\cite{pnud5}.
\begin{figure}[H]
  \centering
    \includegraphics[width=1.\textwidth]{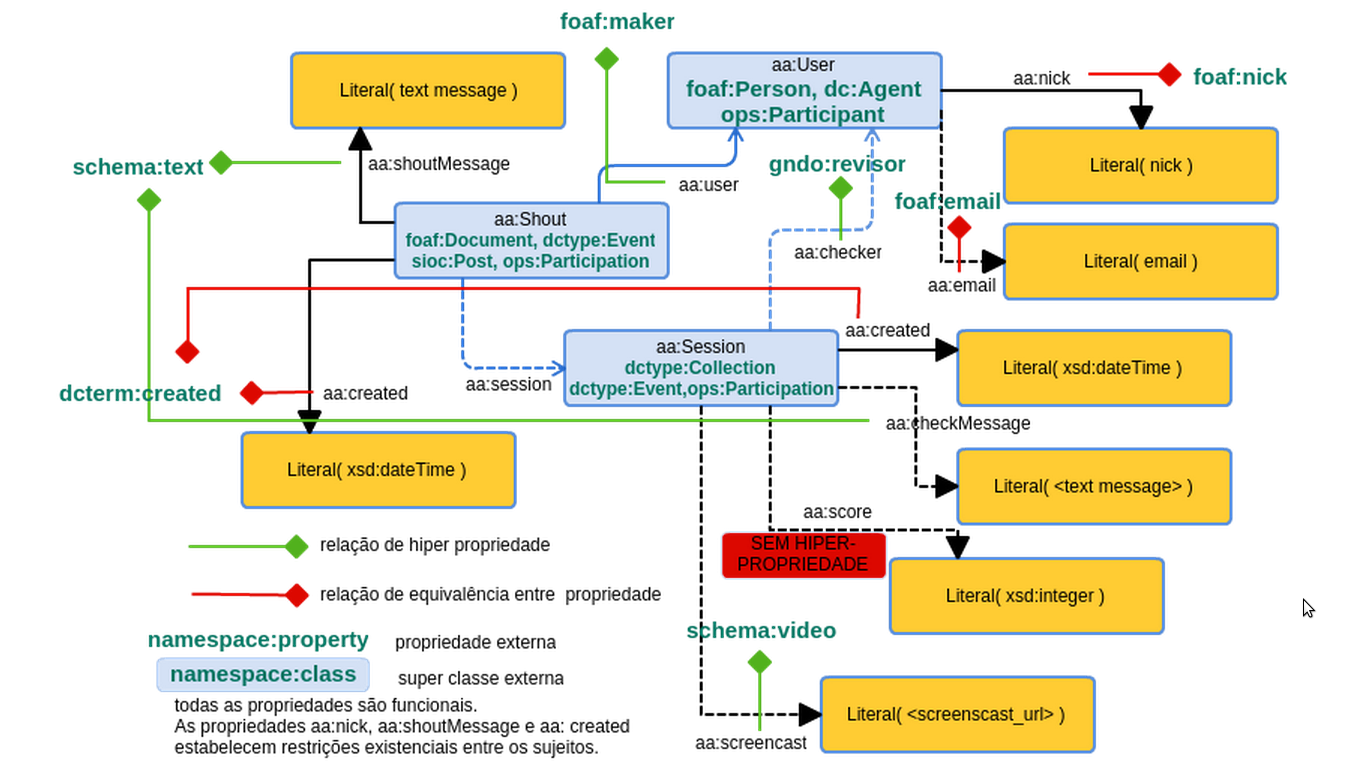}
  \caption{\small Ontologia do AA, com suas classes, propriedades, literais e entidades externas usadas para vincular os dados do AA aos do Participa.br, Cidade Democrática e de toda nuvem LOD.}\label{fig:ontologiaa}
\end{figure}

\begin{figure}[H]
  \centering
    \includegraphics[width=1.\textwidth]{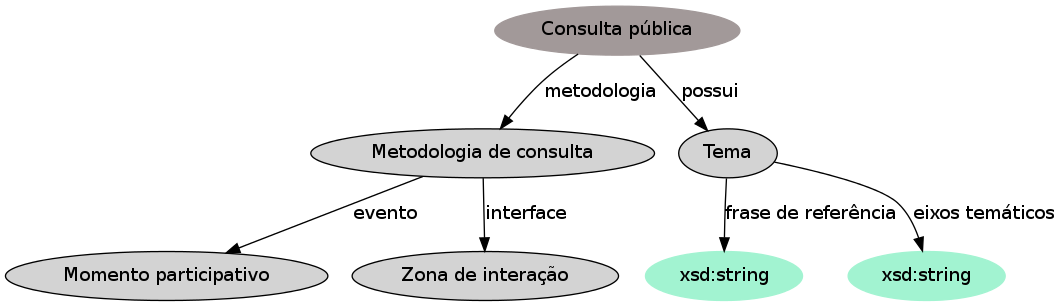}
  \caption{\small Diagrama de uma ontologia simples de consulta pública (ou um embrião de uma ontologia mais complexa). URI passível de derreferenciamento: \url{http://purl.org/socialparticipation/obs/PublicConsultation}, acesse em um browser comum, como Firefox ou Chromium.}\label{fig:consulta}
\end{figure}

A investida maior, porém, foi na Biblioteca (Digital e Semântica de Participação) Social, gerando a OBS (Ontologia da Biblioteca Social) e VBS (Vocabulário da Biblioteca Social). A OBS formaliza em OWL, e a VBS em SKOS, conceitualizações de cada mecanismo e instância de participação social que consta no Decreto 8.243: conferências, conselhos, comissões, ouvidorias, mesas de diálogo, fóruns interconselhos, consultas e audiências públicas. Além disso, contempla documentações produzidas para estas instâncias ou através destas instâncias, e conceitos relevantes, como a PNPS (Política Nacional de Participação Social), SNPS (Sistema Nacional de Participação Social) e a mesa de monitoramento. Os recursos foram publicados no Webprotege da Stanford, em um endpoint SparQL Fuseki/Jena e no Pubby para derreferenciar com redirecionamento do purl.org~\cite{pnud5}. Duas das menores estruturas ontológicas estão nas Figuras~\ref{fig:ontologiaa} e~\ref{fig:consulta}, pois a quantidade de conceitos e relacionamentos implica em uma visualização melhorada quando aberta a figura isoladamente, em PNG.


\section*{Conclusão e próximos passos}
\addcontentsline{toc}{section}{Conclusão e próximos passos}

Direções para nosso aproveitamento estão dadas com os sistemas de recomendação de recursos, seus métodos, polaridades e explicitações~\cite{pnud4}. Como estrutura básica, dados de instâncias participativas estão integrados via critérios semânticos: AA, Cidade Democrática, Participa.br. Ontologias dos mecanismos e instâncias de participação social estão formalizadas em OWL, com vocabulários em SKOS~\cite{pnud5}.

Próximos passos devem incluir a disponibilização para usuários finais a navegação semântica enriquecida com as recomendações via redes complexas e processamento de linguagem natural. Também a integração dos dados linkados/ligados ao ``grafo gigante e global'', da LOD (Linked Open Data), um legado humano de dados conectados.

A etapa atual é de escrita de documentos como este, expondo o que está feito nas áreas da física, da computação, da participação social e das artes. O aprofundamento dos métodos de redes e linguísticos deve se seguir com parceiros do IFSC/USP e acompanhamento da SGPR. Talvez sejam acionadas comunidades relacionadas ao trabalho, como o DIG/MIT ou a W3C, ambos responsáveis por tecnologias e protocolos para os dados ligados/web semântica e utilizados neste trabalho. A simbióse deve continuar com os grupos da UFSCar, IPRJ/UERJ, CCNH/UFABC, UFC, IEA/USP e labMacambira.sf.net, que apoiam estas investidas.

Outras linhas iniciadas e em andamento são:
\begin{itemize}
\item Utilização dos recursos para gerar áudio, música, imagens e videos. Esta linha obteve desenvolvimentos por 3 motivos: 1) para geração de objetos artísticos; 2) enriquecimento da aquisição de informação pela utilização do sentido auditivo junto com a visão; 3) acessibilidade para deficientes visuais.
\item Streaming de estruturas sociais, como nas Figuras~\ref{fig:telao1} e~\ref{fig:telao2}.
\item Integração de dados das diferentes instancias e vinculação à LOD. Este processo, iniciado no AARS, MyNSA e MMISSA~\cite{mynsa,mmissa}, possibilita, por exemplo, buscas textuais e de hashtags no facebook, listas de emails, tweets e outras fontes de informação integradas.
\item Métodos de interação na própria rede. Métodos de ativação da rede como os explicitados no começo deste texto. Métodos de ativação processuais, com iterações ao longo do tempo, e efêmeros, com uma única iteração.
\item Métodos participativos, por exemplo para construção coletiva de textos por etapas e grupos definidos por critérios conectivos (p.ex. periféricos citam substantivos e conceitos principais, hubs qualificam, intermediários montam texto).
\item Gamificação de processos de observação de nossas estruturas em rede, com a navegação, anotação e histórico nos dados ligados.
\item Interfaces para análise de estruturas sociais em evolução temporal, com ênfase em rede e texto .
\end{itemize}

O trabalho já rendeu algumas implementações, ações e documentações de parceiros e terceiros. Exemplos são as citações diretas no produto PNUD dos Profs. Drs. Paulo Meirelles e Fernando Cruz~\cite{paulo6}, na tese de doutorado da Dra. Chandra Wood Viegas~\cite{chandra} e nos objetos artísticos de Pedro Paulo Rocha (Attraktor Zeros)~\cite{pedro}. Este ensaio é o primeiro resumo escrito, para sintonizar os mais imediatamente envolvidos.

\subsection*{Agradecimentos}
\addcontentsline{toc}{section}{Agradecimentos}
Agradeço ao CNPQ (processo 140860/2013-4, projeto 870336/1997-5, orientação: Prof. Dr. Osvaldo Novais de Oliveira Junior); ao PNUD/ONU (contrato 2013/00056, projeto BRA/12/018); aos professores, funcionários, estudantes e comitê de pós-graduação do IFSC/USP; ao DPS/SNAS/SG-PR; aos parceiros do Serpro, UnB, UFABC/CCNH, UFC, IEA/USP, ICMC/USP e labMacambira.sf.net; ao apoio individual dos Profs. Drs. Paulo Meirelles (engenharia de software, UnB), Fernando Cruz (processamento de dados, UnB), Carmen Romcy (consultora PNUD/ONU vinculada à SNAS), Chandra Viegas (IL/UnB), Marília Pisani (CCNH/UFABC), Deborah Antunes (psicologia, UFC), Massimo Canevacci (IEA/USP), Michel Hospital (engenharia elétrica, UFSCar), Ricardo Fabbri (IPRJ/UERJ); aos gestores públicos, pesquisadores, desenvolvedores, gamers, artistas e ativistas Vilson (aut0mata) Vieira, Gabriela Thumé, Daniel (humannoise) Penalva, Danilo (DaneoShiga) Shiga, Edson (prestoppc) Correa, Cássia Wu, Juliana de Souza, Caleb (banzidro) Luporini, Guilherme (cravelho) Lunhani, Guilherme Rebecchi, Glerm e Rodrigo Soares, Simone (Lucida Sans) Bittencourt, Adriano Belisário, Gera (upper) Rocha, Fabiane (Antenna Rush) Borges, George Sanders, Daniel Xavier, Carlos Lobo, Felipe Brait, João Mehl, Iuri Guilherme, Patrícia (angelina) Ferraz, Otávio Martigli, Pedro (Attraktor Zeros) Rocha, Thaís Teixeira, Clovis Souza, Fabiano (contribuição para especificação ontológica de Conferências Nacionais no worshop dia 20/10/2014, sala 101 da SG-PR), Paula Pompeu, Anjuli Osterne (OGU), Paulo Guimarães (OGU), Fernanda Lobato, Valéssio Brito, Silvia (contribuição para especificação ontológica de Consultas Públicas no workshop dia 20/10/2014, sala 101 da SG-PR), Roberto e Márcia (contribuição para especificação ontológica de Mesas de Diálogo no workshop dia 20/10/2014, sala 101 da SG-PR), Henrique Parra, Rodrigo (Yellow) de Luna, Marcos (mquasar) Mendonça, Marcelo Saldanha, Vinícius Rocha, Dina Marques, Mariel Zasso, Fabrício Solagna, Ana Costa, Graziele Machado, Daniela Feitosa, Joênio Costa, Silvio Trida, Ronald Costa, Enaile Iadanza, Lígia Pereira, Pedro Pontual, Gilberto Carvalho e a Excelentíssima Presidenta da República Dilma Rousseff. Ao Ricardo Poppi (diretor geral de novas mídias do Governo Federal) pela supervisão criteriosa dos cinco produtos da consultoria PNUD/ONU resumidos neste texto. Aos Profs. Drs. Leonardo Maia (IFSC/USP), Luciano da Fontoura Costa (IFSC/USP), Osvaldo Novais de Oliveira Junior (IFSC/USP), Dilvan Moreira (ICMC/USP) e Bento Carlos Dias da Silva (CELiC/UNESP) pelo precioso tempo que despreenderam para esta pesquisa, mesmo nos casos breves e indiretos, com reuniões, emails, escrita, reflexão, etc, pois exerceram influência direta no que está sendo feito, o que pretendo retribuir ao menos com textos acadêmicos para eventual publicação.

\bibliographystyle{plain}
\bibliography{ensaio}


\end{document}